\documentclass[a4paper,11pt]{article}
\usepackage{pos}

\title{Results from the KASCADE-Grande Data Analysis}
\ShortTitle{Results from KASCADE-Grande}

\author{D.~ Kang$^{a,*}$,
W.D.~ Apel$^{a}$, J.C.~ Arteaga-Vel\'azquez$^{b}$, K.~ Bekk$^{a}$, M.~ Bertaina$^{c}$,
J.~ Bl\"umer$^{a,d}$, H.~ Bozdog$^{a}$, E.~ Cantoni$^{c,f}$, A.~ Chiavassa$^{c}$,
F.~ Cossavella$^{d}$, K.~ Daumiller$^{a}$, V.~ de Souza$^{g}$, F.~ Di Pierro$^{c}$, P.~ Doll$^{a}$,
R.~ Engel$^{a,d}$, D.~ Fuhrmann$^{h}$, A.~ Gherghel-Lascu$^{e}$, H.J.~ Gils$^{a}$, R.~ Glasstetter$^{h}$,
C.~ Grupen$^{i}$, A.~ Haungs$^{a}$, D.~ Heck$^{a}$, J.R.~ H\"orandel$^{j}$, T.~ Huege$^{a}$,
K.-H.~ Kampert$^{h}$, H.O.~ Klages$^{a}$, K.~ Link$^{a}$, P.~ {\L}uczak$^{k}$, H.J.~ Mathes$^{a}$,
H.J.~ Mayer$^{a}$, J.~ Milke$^{a}$, C.~ Morello$^{f}$, J.~ Oehlschl\"ager$^{a}$,
S.~ Ostapchenko$^{l}$, T.~ Pierog$^{a}$, H.~ Rebel$^{a}$, D.~ Rivera-Rangel$^{b}$, M.~ Roth$^{a}$,
H.~ Schieler$^{a}$, S.~ Schoo$^{a}$, F.G.~ Schr\"oder$^{a}$, O.~ Sima$^{m}$, G.~ Toma$^{e}$,
G.C.~ Trinchero$^{f}$, H.~ Ulrich$^{a}$, A.~ Weindl$^{a}$, J.~ Wochele$^{a}$, J.~ Zabierowski$^{k}$
}
\affiliation[a]{Karlsruhe Institute of Technology, Institute for Astroparticle Physics, Karlsruhe, Germany}
\affiliation[b]{Universidad Michoacana, Inst.~F\'{\i}sica y Matem\'aticas, Morelia, Mexico}
\affiliation[c]{Dipartimento di Fisica, Universit\`a degli Studi di Torino, Italy}
\affiliation[d]{Institut f\"ur Experimentelle Teilchenphysik KIT - Karlsruhe Institute of Technology, Germany}
\affiliation[e]{Horia Hulubei National Institute of Physics and Nuclear Engineering, Bucharest, Romania}
\affiliation[f]{Osservatorio Astrofisico di Torino, INAF Torino, Italy}
\affiliation[g]{Universidade S$\tilde{a}$o Paulo, Instituto de F\'{\i}sica de S\~ao Carlos, Brasil}
\affiliation[h]{Fachbereich Physik, Universit\"at Wuppertal, Germany}
\affiliation[i]{Department of Physics, Siegen University, Germany}
\affiliation[j]{Dept. of Astrophysics, Radboud University Nijmegen, The Netherlands}
\affiliation[k]{National Centre for Nuclear Research, Department of Astrophysics, Lodz, Poland}
\affiliation[l]{Frankfurt Institute for Advanced Studies (FIAS), Frankfurt am Main, Germany}
\affiliation[m]{Department of Physics, University of Bucharest, Bucharest, Romania}

\emailAdd{donghwa.kang@kit.edu}

\abstract{KASCADE-Grande and its original array of KASCADE were dedicated to measure individual air showers of cosmic rays with great detail in the primary energy range of 100 TeV up to 1 EeV. The experiment has significantly contributed to investigations of the energy spectrum and chemical composition of cosmic rays in the transition region from galactic to extragalactic origin of cosmic rays as well as to the further development of hadronic interaction models through validity tests using the multi-detector information from KASCADE-Grande. Though the data accumulation was completed in 2013, the data analysis is still continuing. Recently, we investigate the reliability of the new hadronic interactions model of the Sibyll version 2.3d with the combined data from KASCADE and KASCADE-Grande, and compare it to the predictions of different hadronic interaction models. In addition, we update the web-based platform of the KASCADE Cosmic Ray Data Centre (KCDC), where now full datasets from KASCADE and KASCADE-Grande and the corresponding Monte-Carlo simulated events are available.
}

\FullConference{37$^{\rm{th}}$ International Cosmic Ray Conference (ICRC 2021)\\
		July 12th -- 23rd, 2021\\
		Online -- Berlin, Germany}

\begin{document}
\maketitle

\section{Introduction}
KASCADE \cite{Antoni1} with its extension KASCADE-Grande \cite{Apel1}, measuring individual air showers of cosmic rays, were located at the Karlsruhe Institute of Technology, Karlsruhe, Germany (49.1$^{\circ}$ north, 8.4$^{\circ}$ east, 110\ m a.s.l.).
These experiments measured the energy spectrum, mass composition and the arrival direction of cosmic rays in the primary energy range of PeV to EeV.
The data accumulation was fully completed at the end of 2013 and
all experimental components were dismantled.

KASCADE and KASCADE-Grande reveal important messages
to understand the transition from galactic to extra-galactic cosmic rays:
The all-particle energy spectrum measured by KASCADE shows
a knee-like structure due to a steepening of spectra of light elements \cite{Antoni2}. 
The all-particle energy spectrum reconstructed by the KASCADE-Grande data \cite{Apel2} shows structure,
which does not follow a simple power law:
a concave behavior just above $10^{16}$ eV and a knee-like feature due to heavy primaries, mainly iron, at around $10^{17}$ eV.
In mass composition studies,
the knee-like feature in the heavy primary spectrum is observed more significantly at an energy of 80 PeV \cite{Apel3} in the reconstructed energy spectrum of heavy primary cosmic rays.
In addition, an ankle-like structure is observed at around 100 PeV in the energy spectrum of light primary cosmic rays \cite{Apel4}.

Investigations on the mass composition and energy spectrum of cosmic ray air showers require to understand
high-energy interactions in the Earth's atmosphere,
since the properties of the primary particles by means of extensive air shower
measurements refers from simulations of the air shower development,
in which the hadronic cascade in the atmosphere is described.
Therefore, one of perennial analyses after the completeness of measurements
is the test of hadronic interaction models with KASCADE and KASCADE-Grande data.

Recently, we use full data sets taken by KASCADE-Grande to investigate the validity of a new version of hadronic interaction model Sibyll 2.3d \cite{Sibyll23d}.
In this contribution, we discuss a validity test of Sibyll 2.3d
and its application to the cosmic ray mass composition measured by KASCADE-Grande.

\section{Hadronic interaction model Sibyll 2.3d}
A new version of the hadronic interaction model Sibyll, Sibyll 2.3d, is recently released \cite{Sibyll23d}.
Sibyll is one of event generators for extensive air showers and
it is based on the dual parton and the minijet models.
The interaction cross sections are updated from high-precision measurements of experiments at the LHC
to improve the description of extensive air showers, in particular, the depth of shower maximum and the muon content, 
which affect the interpretation of the mass composition of the primary cosmic rays.
Regarding the impact on extensive air showers, 
the muon number in the updated model increased by more than 20\% compared to Sibyll 2.1,
since it predicted too less muons.

Compared to the other interaction models EPOS-LHC and QGSjet-II-04, Sibyll 2.3d has a higher number
of muons, but only by about 5\%.
Further details can be found in Ref. \cite{Sibyll23d}.

For the air shower simulations the program CORSIKA \cite{Heck}
has been used,
where the FLUKA (E $<$ 200\ GeV) model has been used for hadronic interactions at low energies
and high-energy interactions were treated with Sibyll 2.3d.
Showers initiated by five different primaries (H, He, CNO, Si, and Fe nuclei)
have been simulated. The
simulations covered the energy range of 10$^{14}$ - 10$^{18}$ eV with zenith
angles in the interval 0$^{\circ}$ - 42$^{\circ}$. 
For the analysis we used zenith angles only up to 40$^{\circ}$, where full efficiency is reached.
The spectral index in the simulations was -2 and for the analysis it is converted to a slope of -3.

\section{Data analysis}

\subsection{Shower size}
Using the shower size measured by KASCADE-Grande data, initial tests of Sibyll 2.3d were performed.
Figure\ 1 (left) shows the 2-dimensional shower size measured by KASCADE-Grande, including the full detector response by simulation, together with proton and iron induced showers
predicted by different interaction models: Sibyll 2.3d (red), Sibyll 2.3 (blue), EPOS-LHC (green) and QGSjet-II-04 (pink).
The errors of mean values are plotted in Fig.\ 1.
Sibyll 2.3d has a similar tendency compared to the other interaction models above the threshold ($N_{\mu} \gtrsim 5$, $N_{ch} \gtrsim 6$), whereas it shows some deviations below the threshold for iron showers.

On the right panel of Fig.\ 1 residual plots for simulations compared to different interaction models
with respect to Sibyll 2.3d can be seen.
The top plot is for iron shower and the bottom for proton.
QGSjet-II-04 and Sibyll 2.3 present a similar deviation of about 30\% relative to Sibyll 2.3d.
EPOS-LHC shows the smallest difference from Sibyll 2.3d.
In addition, EPOS-LHC has more statistics than other models by a factor of 1.5, so that smaller statistical error bars are shown.

\begin{figure}[t!]
  \begin{center}
    \includegraphics[width=0.52\textwidth]{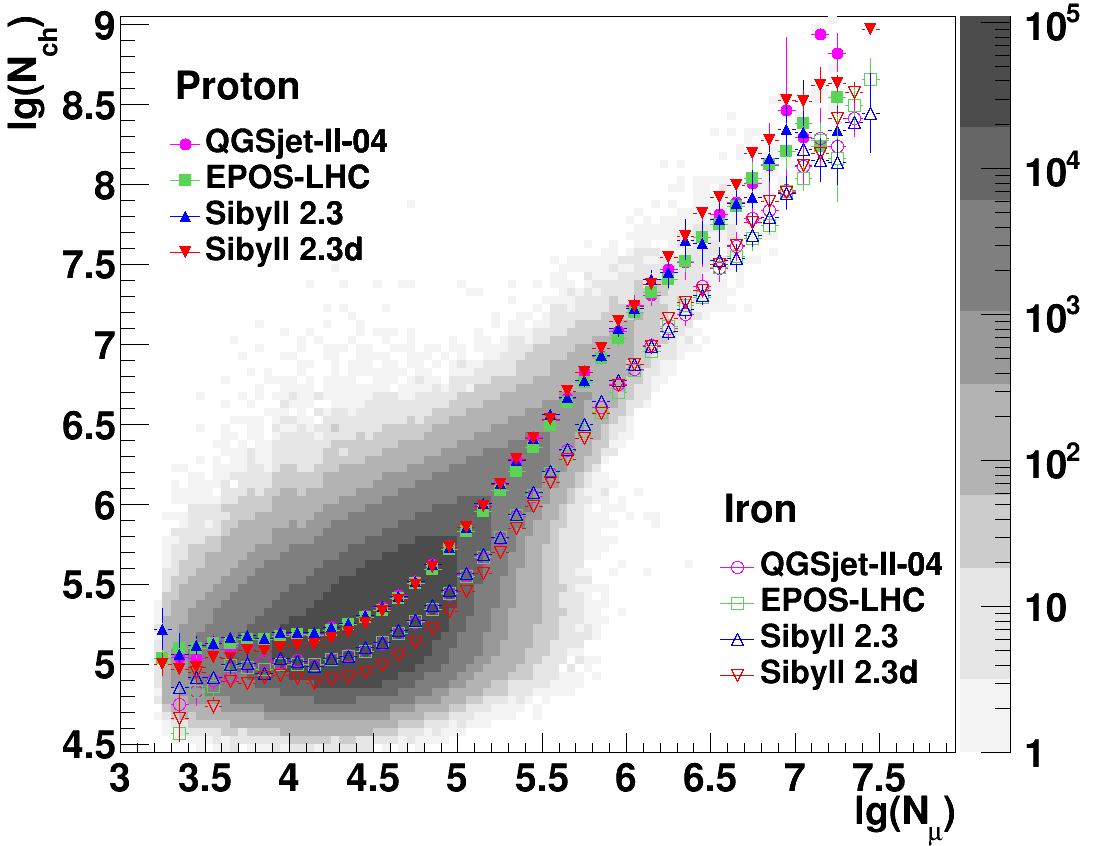}
    \includegraphics[width=0.47\textwidth]{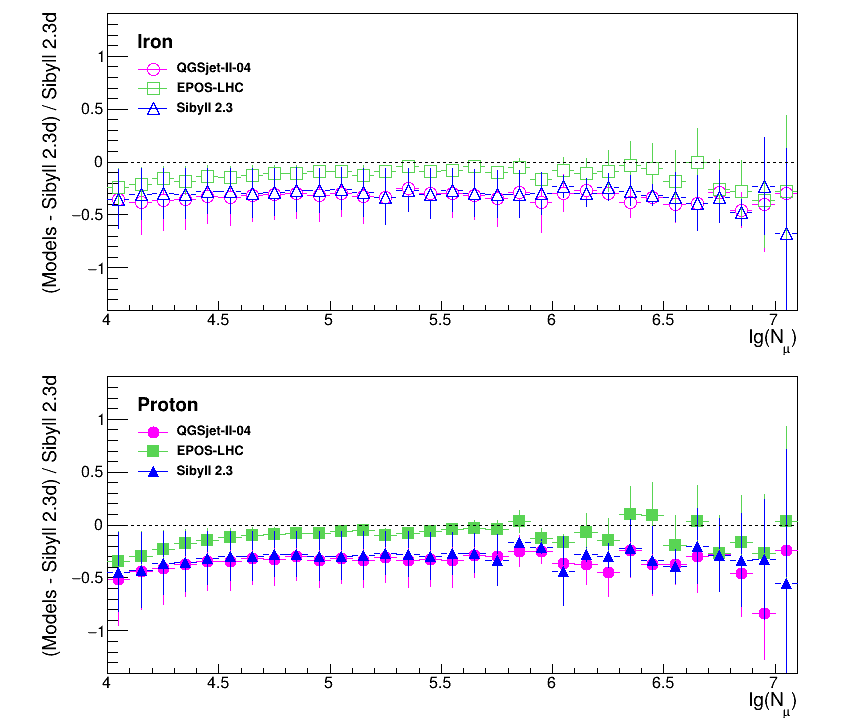}
    \caption{Left: The 2-dimensional shower size distribution measured by KASCADE-Grande, overlaying proton and iron induced showers for different hadronic interaction models.
    Right: Residual plots for simulations (proton and iron) comparing three different models with respect to Sibyll 2.3d.}
\label{fig1}
\end{center}
\end{figure}

\begin{figure}[t!]
  \begin{center}
    \includegraphics[width=0.49\textwidth]{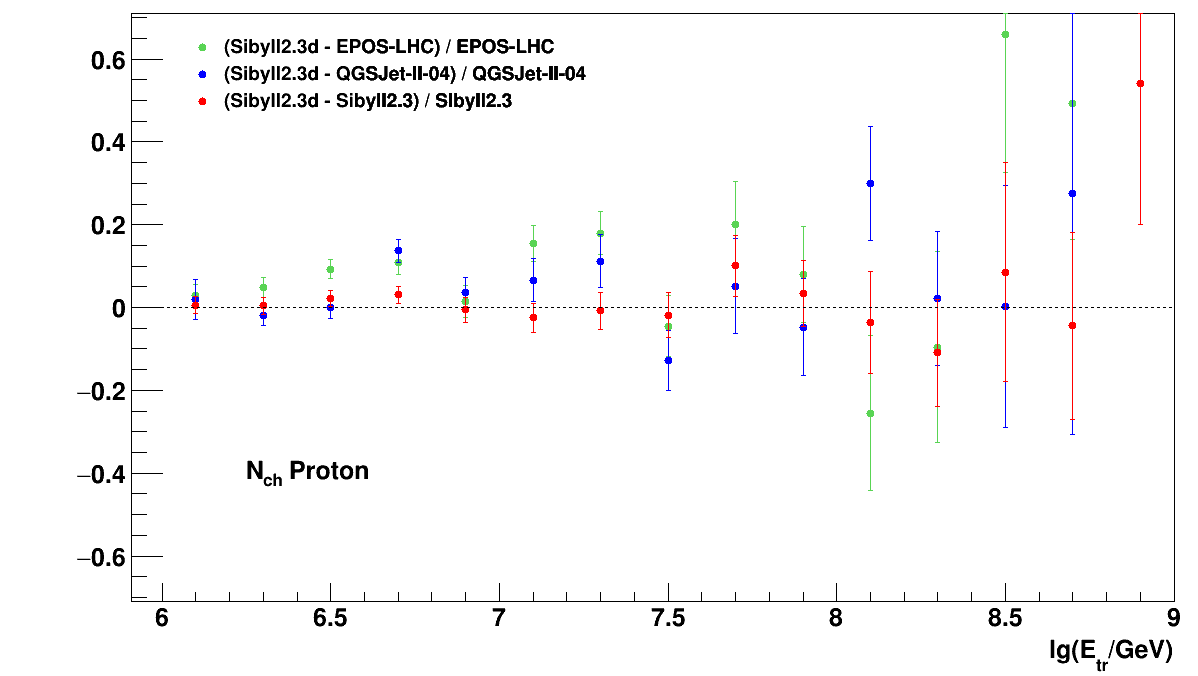}
    \includegraphics[width=0.49\textwidth]{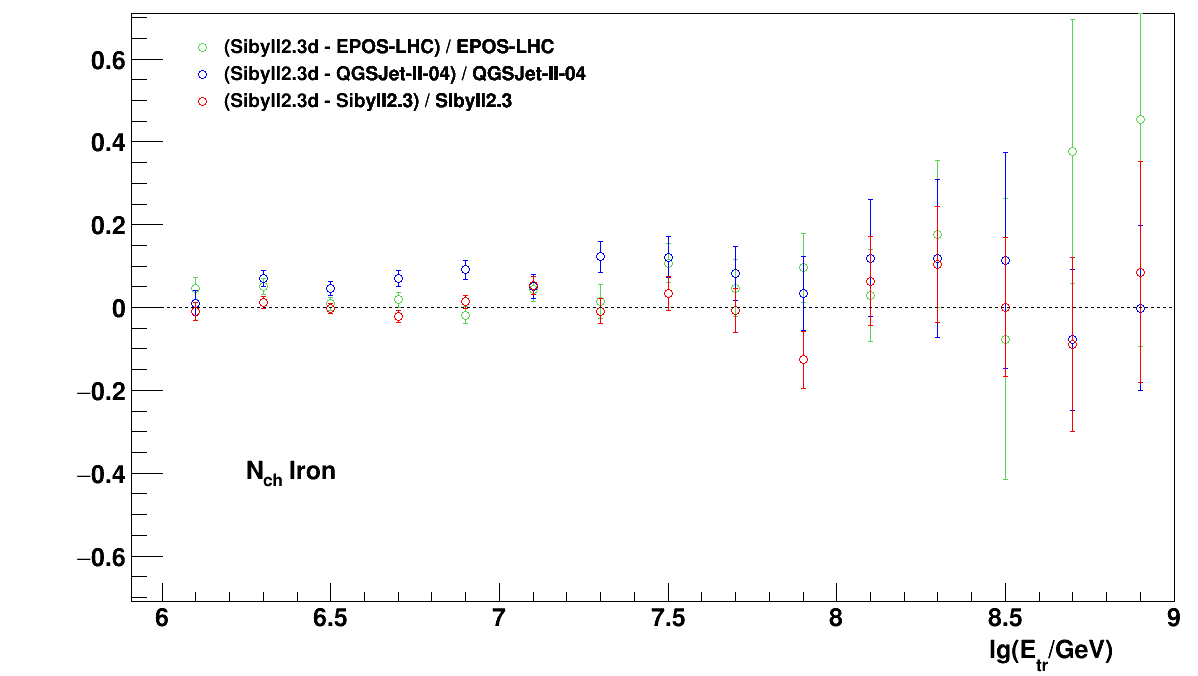}
    \includegraphics[width=0.49\textwidth]{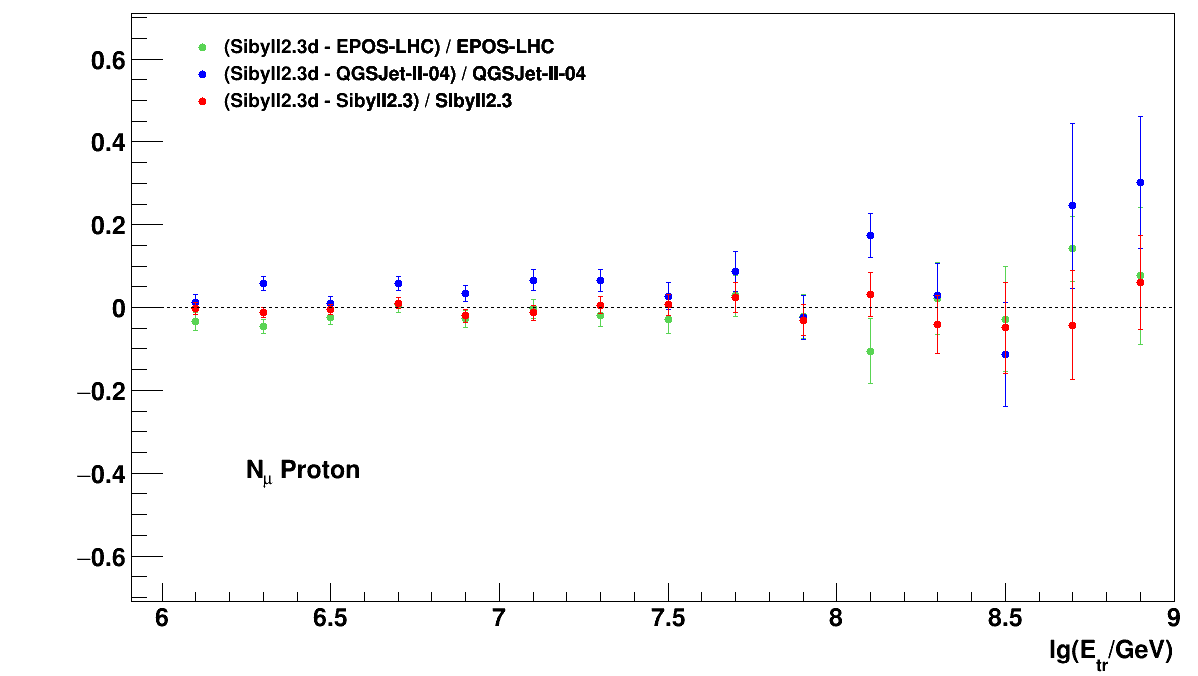}
    \includegraphics[width=0.49\textwidth]{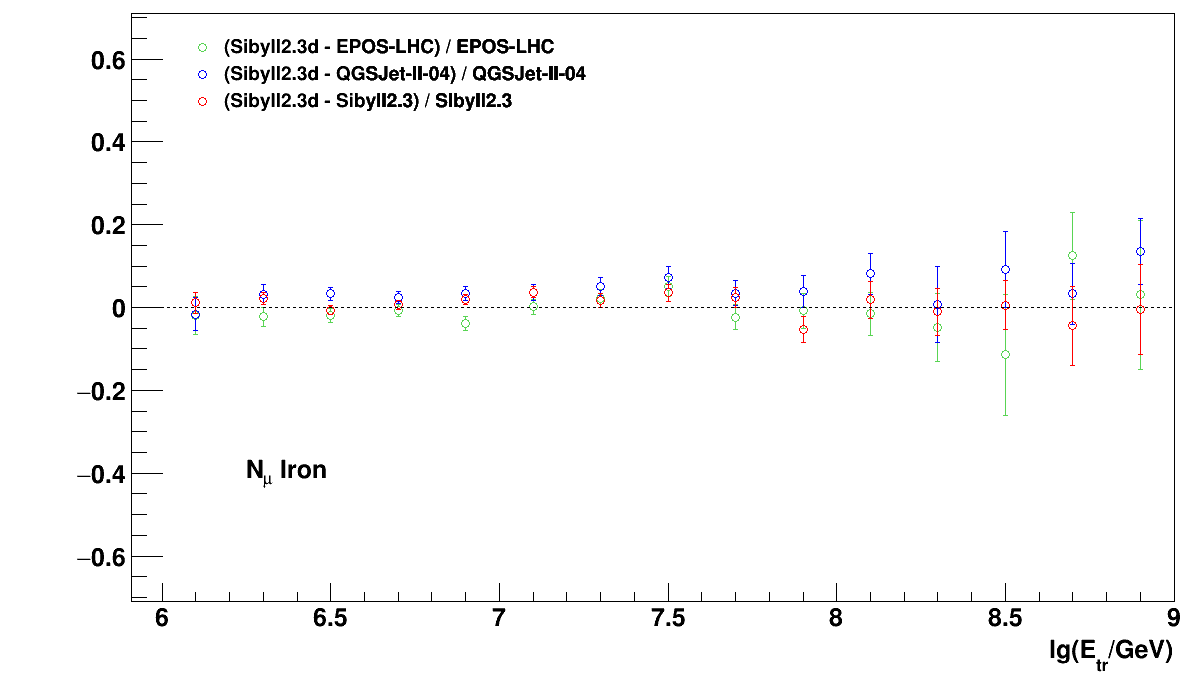}
    \caption{Comparison of the number of charged particles ($N_{ch}$) and the number of muons ($N_{\mu}$) of Sibyll 2.3d to other previous interaction models for proton (left) and iron (right) primaries, respectively.}
\label{fig2}
\end{center}
\end{figure}
Figure\ 2 presents comparisons of the number of charged particles ($N_{ch}$) and the number of muons ($N_{\mu}$) of Sibyll 2.3d to other previous interaction models. The upper plots are $N_{ch}$ and the lower $N_{\mu}$.
The iron induced showers fluctuate less than the proton showers.
Sibyll 2.3 shows the smallest difference from Sibyll 2.3d for all cases.
The largest difference shows the case of QGSjet-II-04, however it is about the level of 5\% differences.

\subsection{Separation into mass groups}
In this analysis electron-rich and -poor samples are separated by using the shower size ratio of $Y_{CIC} = lg(N_{\mu})/lg(N_{ch})$,
where the Constant Intensity Cut (CIC) technique is used to correct the muon
and charged particle numbers for attenuation effects in the atmosphere.
The events which satisfy the condition ($Y_{CIC} \geq Y^{thr}_{CIC}$)
are defined as electron-poor events and the remaining ones as electron-rich events.
The dotted line in Fig.\ 3 represents the selection criteria of $Y^{thr}_{CIC}$.
The value is model dependent and it is defined to be between the silicon and the CNO element for each interaction model.

\begin{figure}[t!]
  \begin{center}
    \includegraphics[width=0.55\textwidth]{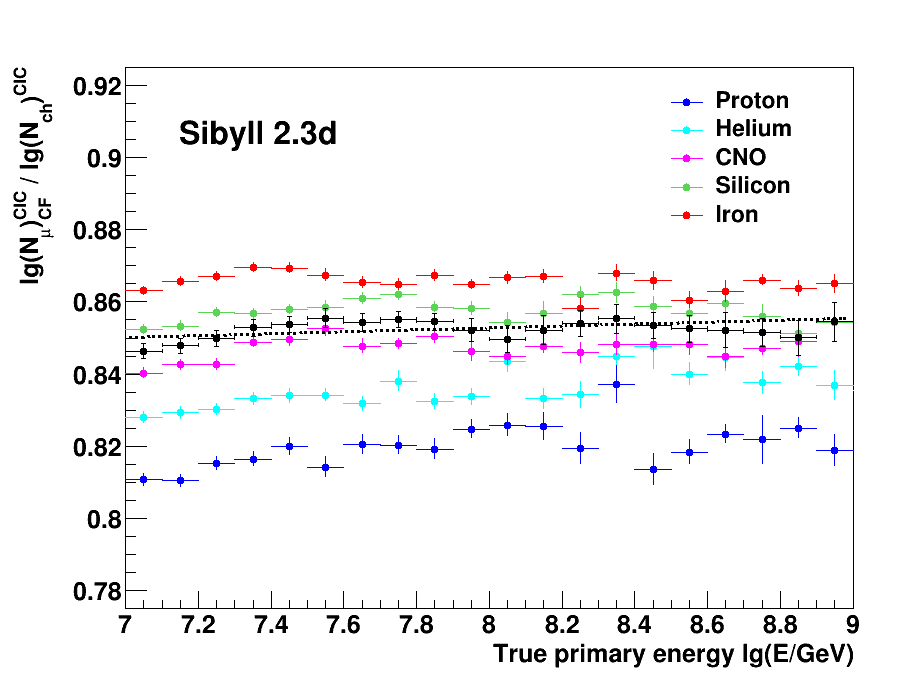}
    \caption{Shower size ratio of $Y_{CIC} = lg(N_{\mu})/lg(N_{ch})$ as a function of true energy,
      where $N_{\mu}$ and $N_{ch}$ are corrected for attenuation effects in atmosphere using the CIC technique.
      The dotted line is the separation criteria for heavy and light mass groups.
    } 
\label{fig3}
\end{center}
\end{figure}

\subsection{Energy calibration}
The energy calibration function for light and heavy induced showers is shown in Fig.\ 4 (left).
On the right of Fig.\ 4 the comparison with the different versions of Sibyll is shown.
Under the assumption of a linear dependence in logarithmic scale:
lg$E$ = $a\cdot$lg($N_{ch}$) + $b$ and a particular primary composition,
the fitting is applied in the range of full trigger and reconstruction
efficiencies.
The energy calibration depends on simulations, i.e. interaction models,
so that the fits are performed individually and
the resulting coefficients of the calibration for Sibyll 2.3d are $a = 0.891 \pm 0.004$, $b = 1.802 \pm 0.024$ and $a = 0.943 \pm 0.005$, $b = 1.216 \pm 0.035$ for heavy and light primaries, respectively.
The slope of Sibyll 2.3d is quite similar to the one of Sibyll 2.3.
Using this fit function, we converted the attenuation corrected shower size into the reconstructed energy.

\begin{figure}[t!]
  \begin{center}
    \includegraphics[width=0.47\textwidth]{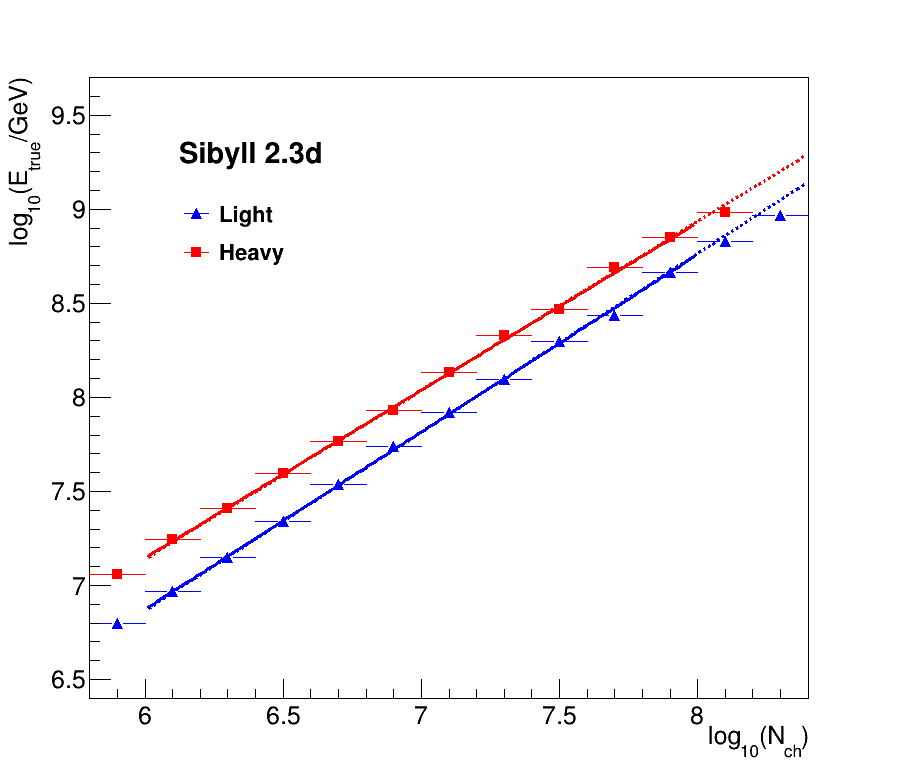}
    \includegraphics[width=0.51\textwidth]{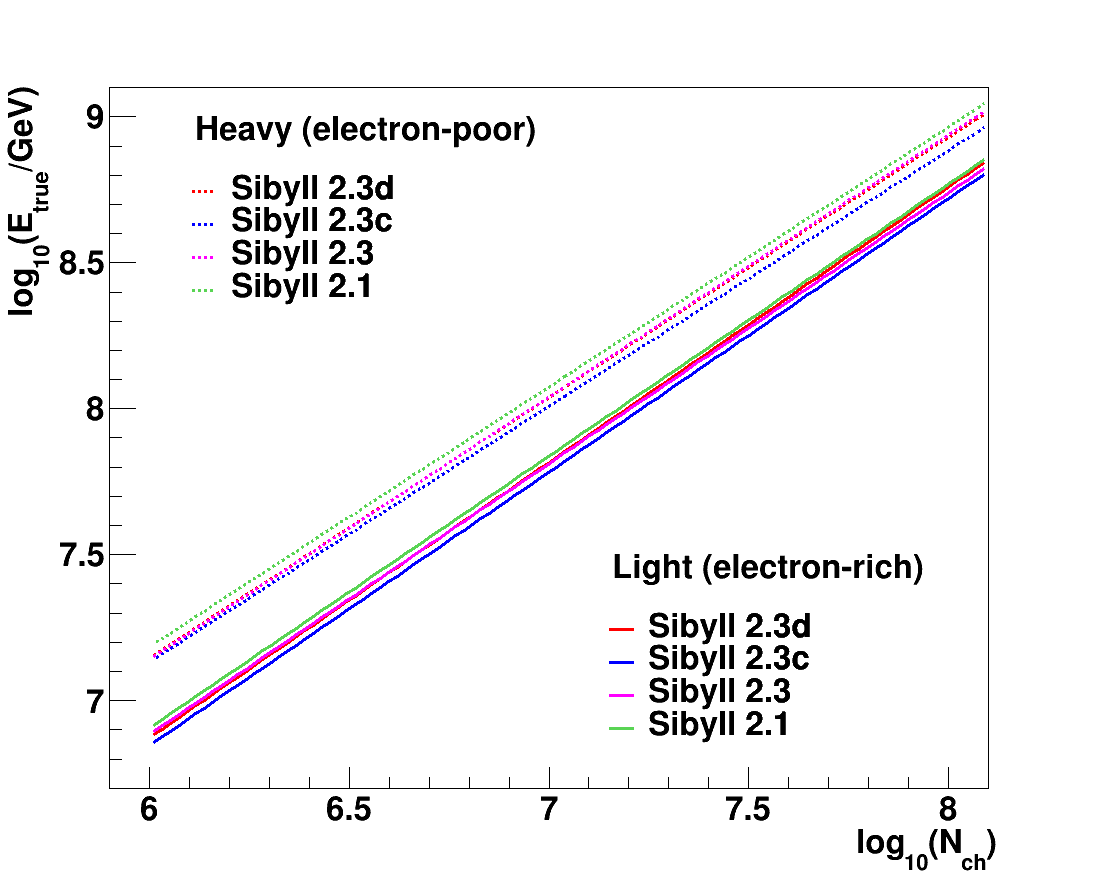}
    \caption{Left: The true energy as a function of the number of charged particles ($N_{ch}$) for light and heavy primaries for Sibyll 2.3d. Right: Comparison of the energy calibration function of light and heavy primaries for previous Sibyll models.
 }
\label{fig4}
\end{center}
\end{figure}

\section{Spectra of heavy and light mass groups}
The energy is assigned using the relation $E(N_{ch})$ for the two separated samples,
using calibration functions which are model dependent.

\begin{figure}[t!]
  \begin{center}
    \includegraphics[width=0.55\textwidth]{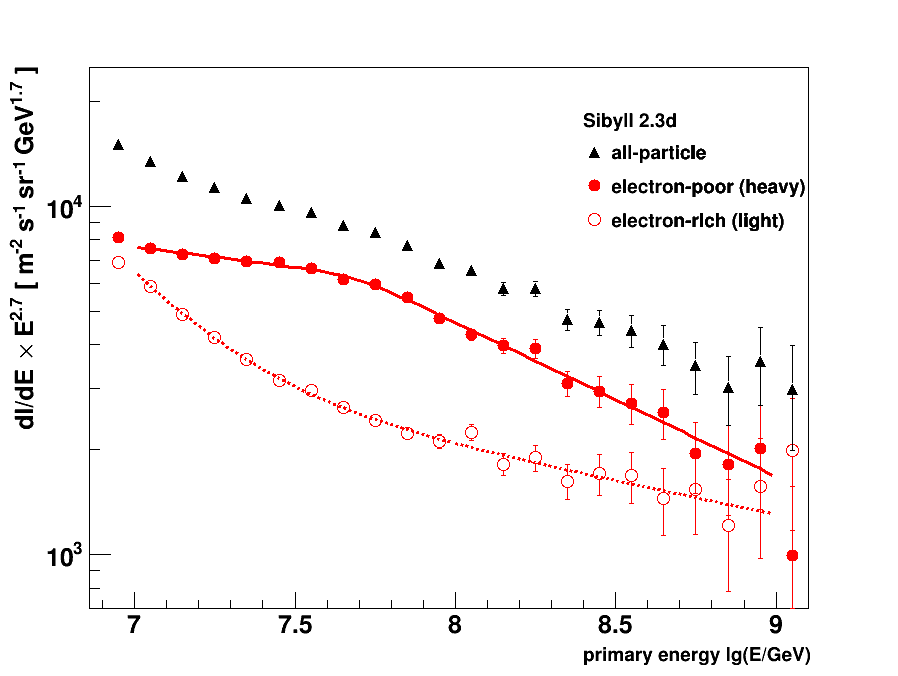}
    \caption{
      The resulting energy spectra of heavy and light primaries based on the Sibyll 2.3d model,
      fitting with a broken power law.}
  \label{fig5}
  \end{center}
\end{figure}

\begin{figure}[t!]
  \begin{center}
    \includegraphics[width=0.49\textwidth]{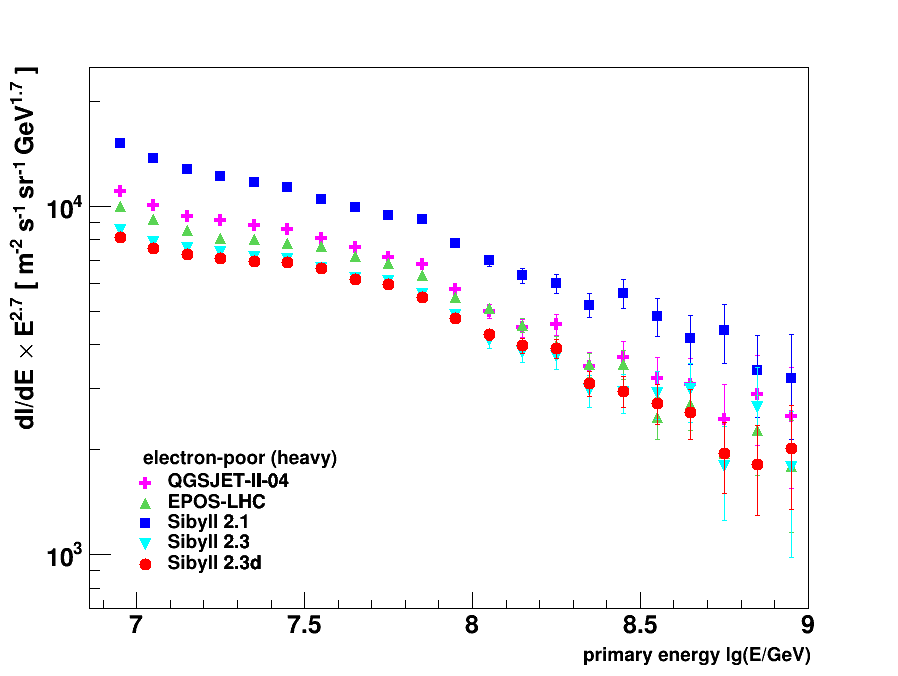}
    \includegraphics[width=0.49\textwidth]{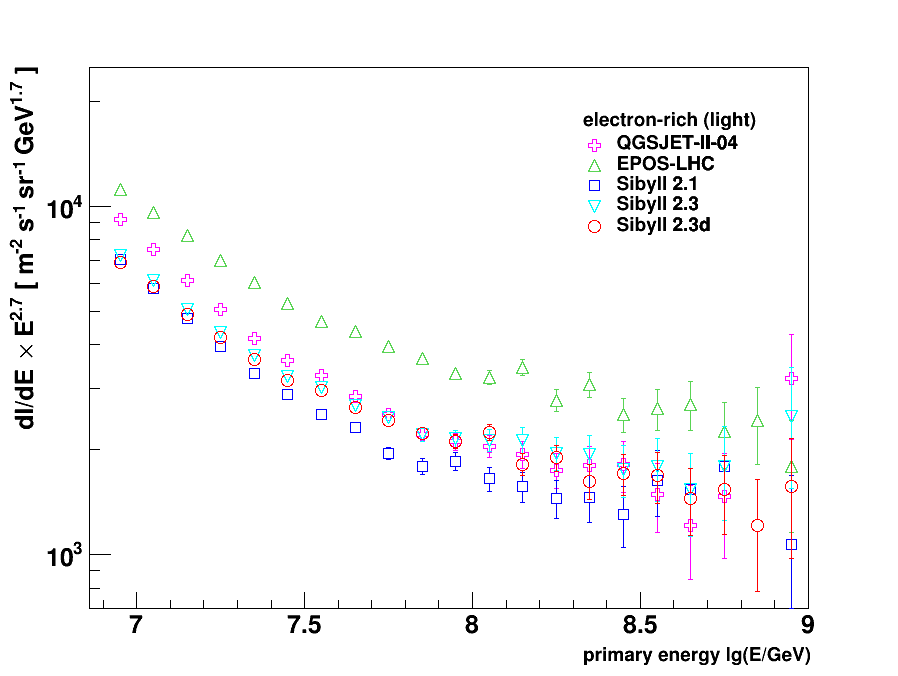}
    \caption{
      Comparisons of the reconstructed energy spectra for four hadronic interaction models.
      The left plot is for the heavy mass group and the right one is for light primaries. 
      The error bars show the statistical uncertainties.}
  \label{fig6}
  \end{center}
\end{figure}

Figure 5 shows the resulting reconstructed energy spectra of heavy and light mass group based on the Sibyll 2.3d model, with only statistical errors.
Systematic uncertainties are under investigation, however, they are expected to be about 25\% on the flux. 
The fit of a broken power law has been performed for the spectra.
Resulting slopes before and after the heavy knee and the breaking positions are indicated in Table 1.
All features observed by the previous analysis are well confirmed.
The spectrum of heavy primaries, i.e. electron-poor events, shows a clear knee-like structures at $10^{16.7}$ eV.
A remarkable feature is that the concave structure 
at about $10^{16}$ eV is more visible in the spectrum of electron-poor components.
In the light primary spectrum, the spectral slope changes smoothly and a hardening feature above about $10^{16.5}$ eV is observed.

Reconstructed energy spectra of the electron-poor and electron-rich groups,
based on different post-LHC models of QGSjet-II-04, EPOS-LHC, Sibyll 2.1, Sibyll 2.3 and Sibyll 2.3d,
were compared in Fig.\ 6,
in which all spectra were reconstructed by applying the Constant Intensity Cut technique.
All spectra shown in Fig 6 are not corrected yet for shower fluctuations.

In the comparison of the energy spectra based on the Sibyll 2.1 model,
the flux of heavy primaries of Sibyll 2.3d shows differences by a factor of about 3. 
The total flux is shifted about 10-20\%, but the general structure are similar.
The muon content might affect the difference of absolute abundances and detailed studies can be found in Ref. \cite{JuanCarlos}.

\begin{table}
\begin{center}
\begin{tabular}{lccccc}
\hline
electron-poor & $lg(E_{k}/GeV)$ & $\gamma_{1}$  & $\gamma_{2}$   & $\Delta \gamma$ & $\chi^{2}$/ndf \\ \hline
QGSjet-II-04  & 7.73 $\pm$ 0.05  & 2.89 $\pm$ 0.01 & 3.18 $\pm$ 0.04 & 0.29            & 2.16 \\
EPOS-LHC      & 7.79 $\pm$ 0.03  & 2.87 $\pm$ 0.01 & 3.20 $\pm$ 0.03 & 0.33            & 4.72 \\
Sibyll 2.1    & 7.75 $\pm$ 0.09  & 2.87 $\pm$ 0.03 & 3.15 $\pm$ 0.05 & 0.28            & 1.28 \\
Sibyll 2.3    & 7.71 $\pm$ 0.05  & 2.83 $\pm$ 0.01 & 3.18 $\pm$ 0.05 & 0.35            & 0.96 \\
Sibyll 2.3d   & 7.69 $\pm$ 0.04  & 2.82 $\pm$ 0.01 & 3.14 $\pm$ 0.03 & 0.32            & 1.47 \\ \hline
\end{tabular}
\caption{\label{label}The breaking positions and the spectral slopes after applying a broken power law fit to the spectra of electron-poor events}
\end{center}
\end{table}

\section{Conclusion}
Based on the new hadronic interaction model Sibyll 2.3d and the shower size measured by KASCADE-Grande,
the energy spectra of different mass groups were reconstructed.
It was compared with the spectra based on the different post-LHC models.
All features of the energy spectra confirmed by previous measurements are shown:
observation of a heavy knee at around $10^{17}$ eV and
flattening of the light component at about $10^{17}$ eV.
This might be a sign of an extra-galactic component and it is already dominant below the energy of $10^{17}$ eV for the case of Sibyll 2.3d model.

According to the comparison of the shower size of the new model Sibyll 2.3d,
it is observed that Sibyll 2.3d has a higher number of muons compared to other models as this model expected.
In addition, this model gives the lowest flux of heavy primaries of all models, i.e. the lightest composition.
Detailed studies including estimation of systematic uncertainties and the correction of shower fluctuations
are in progress.

Lastly, the full experimental data sets and the simulations with detector responses can be found in KASCADE Cosmic Ray Data Centre (KCDC) \cite{KCDC}, which is a pioneering work in public access of astroparticle physics data.


\begin{thebibliography}{99}
\bibitem{Antoni1} T. Antoni et al., KASCADE Coll., Nucl. Instr. Meth. A513 (2003) 490 
\bibitem{Apel1} W.D. Apel et al., KASCADE-Grande Coll.,  Nucl. Instr. and Meth. A620 (2010) 202
\bibitem{Antoni2} T. Antoni et al., KASCADE Coll., Astropart. Phys. 24 (2005) 1-25 
\bibitem{Apel2} W.D. Apel et al., KASCADE-Grande Coll., Astrop. Phys. 36 (2012) 183 
\bibitem{Apel3} W.D. Apel et al., KASCADE-Grande Coll., Phys. Rev. Lett. 107 (2011) 171104 
\bibitem{Apel4} W.D. Apel et al., KASCADE-Grande Coll., Phys. Rev. D 87 (2013) 081101 
\bibitem{Sibyll23d} F. Riehn et al., Phys. Rev. D 102 (2020) 063002 
\bibitem{Heck} D. Heck et al., Report Forschungszentrum Karlsruhe, FZKA 6019 (1998)
\bibitem{JuanCarlos} J.C. Arteaga-Vel\'azquez et al., KASCADE-Grande Coll., PoS(ICRC2021)376, these proceedings 
\bibitem{KCDC} A. Haungs et al., PoS(ICRC2021)422, these proceedings  
\end{thebibliography}
\end{document}